\documentclass[journal]{IEEEtran}
\long\def\/*#1*/{}
\usepackage{amsmath,amsfonts}
\usepackage{lmodern}
\usepackage{graphicx}
\usepackage{caption}
\usepackage{subcaption}
\begin{document}
\title{Entropic Empirical Mode Decomposition}
\author{
    \IEEEauthorblockN{Sumit Kumar Ram\IEEEauthorrefmark{1}, Marta Molinas\IEEEauthorrefmark{2}}
\\   \IEEEauthorblockA{\IEEEauthorrefmark{1}\IEEEauthorrefmark{2}Dept. of Eng. Cybern., Norwegian Univ. of Sci. and Tech., Trondheim, Norway.
\\    \IEEEauthorblockA{\IEEEauthorrefmark{1}Indian Institute of Science Education and Research, Kolkata, Mohanpur, Nadia, India
    \\\IEEEauthorrefmark{1}\{skr12ms039\}@gmail.com}
     \\\IEEEauthorrefmark{2}\{marta.molinas\}@ntnu.no}
}
\maketitle

%%%%%%%%%%%%%%%%%%%%%%%%%%%%%%%%%%%%%%%%%%%%%%%%%%%%%%%%%%%%%%%%%%%%%%%%%%%%%%%%%%%%%%%%%%%%%%%%%%%%%%%%%%%%%%%%%%%%%%%%%%%%%%

%%%%%%%%%%%%%%%%%%%%%%%%%%%%%%%%%%%%%%%%%%%%%%%%%%%%%%%%%%%%%%%%%%%%%%%%%%%%%%%%%%%%%%%%%%%%%%%%%%%%%%%%%%%%%%%%%%%%%%%%%%%%%%
\begin{abstract}
\begin{quote}
Empirical Mode Decomposition(EMD) is an adaptive data analysis technique for analyzing nonlinear and non stationary data\cite{huang1}. EMD decomposes the original data into a number of Intrinsic Mode Functions(IMFs)\cite{huang1} for giving better physical insight of the data. Permutation Entropy(PE) is a complexity measure\cite{bandt} function which is widely used in the field of complexity theory for analyzing the local complexity of time series. In this paper we are combining the concepts of PE and EMD to resolve the mode mixing problem observed in determination of IMFs.    

\end{quote}
\end{abstract}
%%%%%%%%%%%%%%%%%%%%%%%%%%%%%%%%%%%%%%%%%%%%%%%%%%%%%%%%%%%%%%%%%%%%%%%%%%%%%%%%%%%%%%%%%%%%%%%%%%%%%%%%%%%%%%%%%%%%%%%%%%%%%%

%%%%%%%%%%%%%%%%%%%%%%%%%%%%%%%%%%%%%%%%%%%%%%%%%%%%%%%%%%%%%%%%%%%%%%%%%%%%%%%%%%%%%%%%%%%%%%%%%%%%%%%%%%%%%%%%%%%%%%%%%%%%%%
\section{Introduction}
 Empirical Mode Decomposition proposed by Huang Et al.\cite{huang1} serves as a widely used and effective method for analyzing nonlinear and non stationary data\cite{huang3}. It decomposes the original data into a finite and often small number of well behaved Hilbert transformable\cite{huang4} intrinsic mode functions(IMFs). Due to its adaptive nature and instantaneous frequency\cite{huang4} determinable IMFs it gives a sharp identification of the embedded structure and hence is highly efficient in terms of analyzing physical reality\cite{huang3} of the data.\par
  "Mode mixing" is defined as any IMF consisting of oscillations of dramatically disparate scales\cite{huang2}, mostly caused by intermittency of the driving mechanics. Mode mixing leads to deceptive physical meaning of the mixed mode IMFs\cite{huang3}. Hence mode mixing problem is the major drawback of the EMD and was first illustrated by Huang et al.\cite{huang3}\par
  Permutation entropy proposed by Bandt et al\cite{bandt} is an effective way of measuring the complexity of the time series by comparing the neighboring values around each point. Further Cao et al.\cite{cao} showed that it is an efficient way for Detecting dynamical changes in time series. This shows the adaptive nature\cite{liu} of the technique. Hence it can be concluded that the combination of EMD and PE will maintain its adaptivity and can be used for adaptive data analysis.
%%%%%%%%%%%%%%%%%%%%%%%%%%%%%%%%%%%%%%%%%%%%%%%%%%%%%%%%%%%%%%%%%%%%%%%%%%%%%%%%%%%%%%%%%%%%%%%%%%%%%%%%%%%%%%%%%%%%%%%%%%%%%%

%%%%%%%%%%%%%%%%%%%%%%%%%%%%%%%%%%%%%%%%%%%%%%%%%%%%%%%%%%%%%%%%%%%%%%%%%%%%%%%%%%%%%%%%%%%%%%%%%%%%%%%%%%%%%%%%%%%%%%%%%%%%%%
\section{Empirical Mode Decomposition}

 Empirical Mode Decomposition proposed by Huang et al. (1998)\cite{huang1} is an effective algorithm to decompose time series into the so-called Intrinsic Mode Functions. These modes form a (quasi-)orthogonal set of basis functions that is derived directly from the original data without a priory assumptions\cite{huang1} about their nature. In addition to orthogonality of the total set of basis functions, each IMF should have the following property

 \begin{itemize}
 \item Number of extreme values and zero crossings differ at most by one.
 \item The "local mean" value of the mode is zero .
 \end{itemize}

In order to get the IMFs from the original data set we can follow the following algorithm

\begin{enumerate}
\item Start with a discrete time series x(t) with sampling period Ts.

\item Compute all maximum (minimum) values of x(t), perform cubic
spline interpolation, $s_{max} (s_{min})$, for these values.

\item Estimate the local mean of the time series as 
$$m(t)=\frac{1}{2}(s_{max}+s_{min})$$

\item Subtract m(t) from x(t) and repeat $step- 1,2,3$ with the resulting
time series, until a stopping criterion is reached.

\item Once the stopping criterion is reached name the function as first IMF ($c_1$). Subtract this IMF from the original signal and repeat the $steps-1,2,3,4$ until getting the residue.

\item Residue is characterized as the mode function which does not have enough extremas for interpolation. This residual can be regarded as the global trend, r(t), of the time series
\end{enumerate}
   Hence with the EMD algorithm we can decompose any time series into its IMFs($c_i$) and residue($r(t)$) as 
   $$x(t)=\sum_{i=1}^{n} c_i(t)+r(t) $$
   where n is the total number of IMFs obtained from the original time series.

%%%%%%%%%%%%%%%%%%%%%%%%%%%%%%%%%%%%%%%%%%%%%%%%%%%%%%%%%%%%%%%%%%%%%%%%%%%%%%%%%%%%%%%%%%%%%%%%%%%%%%%%%%%%%%%%%%%%%%%%%%%%%%

%%%%%%%%%%%%%%%%%%%%%%%%%%%%%%%%%%%%%%%%%%%%%%%%%%%%%%%%%%%%%%%%%%%%%%%%%%%%%%%%%%%%%%%%%%%%%%%%%%%%%%%%%%%%%%%%%%%%%%%%%%%%%%
\section{Permutation Entropy}
The quantification of complexity of a time series plays a crucial role in biomedical analysis\cite{zanin} as well as in other branches of nonlinear sciences. Various complexity measures have been proposed and is used to study the complexity of time series data to get the useful information about the temporal dynamical changes in time series. Some of the widely used complexity measure are correlation dimension and the Kolmogorov-Sinai entropy and Permutation entropy (PE). Permutation Entropy is a conceptually simple method that allows the analysis and representation of a time series through a symbolic sequence. The advantages of the PE method reside in its simplicity, its robustness\cite{bandt}. In addition, this method provides an extremely fast computational algorithm and it can be applied to any type of time
series.\par
For calculating the PE of a given time series we can proceed with the following procedure.\par
  Consider the measured scalar time series data $x(t_i)$, for n = 1, 2,..., where i denotes the i-th variable of the system.
 Firstly, we divide the long time series into segments with fixed finite length $\tau$(for convenience we have chosen the length $\tau$ to be 120 here) which can be denoted as $\tau_j$ (j = 1, 2,....). These $\tau_js$ are obtained by applying running windows over our original time series. In our case we have chosen the step size $1$ to have a better information about the complexity variation over time. Further we use another sliding-window analysis to partition each $\tau_j$ into short sequences of given length m. For convenience we have taken the value of m to be 3 here. Fig. 1 describes the complete procedure of sliding windows graphically.
  In this we for a segment of time series with length $\tau_i$ , the total number of short sequences will be $\tau_i-m+1$. Inside a short sequence, we distinguish it as a pattern by the natural position order. It can be easyly observed that for our case $m = 3$, we can have six different possibilities such as
\begin{itemize}
 \item $x(t_i) < x(t_{i+1})$
 \item $x(t_i) < x(t_{i+2})$,
 \item $x(t_{i+1})<x(t_{i})$
 \item $x(t_{i+1})<x(t_{i+2})$,
 \item $x(t_{i+2})<x(t_{i})$
 \item $x(t_{i+2})<x(t_{i+1})$
\end{itemize}
 Using $p_i$ $(i = 1,..., 6)$ to denote the probability that one of the patterns appears in the time series of a segment, we can define the permutation entropy $H(\tau_j)$ as
     $$H(\tau_j)=-\sum_{i=1}^{3!}p_iln(p_i) $$

	For making $p_i$ roughly reflect the local structure of time series, the number of short sequences in $\tau$ should not be too small, i.e.,$N_i-m+1>100$\cite{liu}.
	
 Because the time series is always changing with time, $H(\tau_j)$ changes with time period interval $\tau_j$ and will be determined by the local topological structure of the time series\cite{liu}. For determining the complexity measure in terms of permutation entropy\cite{bandt} around each point of our time series we are calculating the permutation entropy by taking the running windows of $\tau_i$ along the whole time series data.
%\end{Permutation Entrpy Calculation}
\begin{figure}	
	\centering
\begin{subfigure}[t]{1in}
\centering
\hspace*{-3cm}
\includegraphics[scale=0.38]{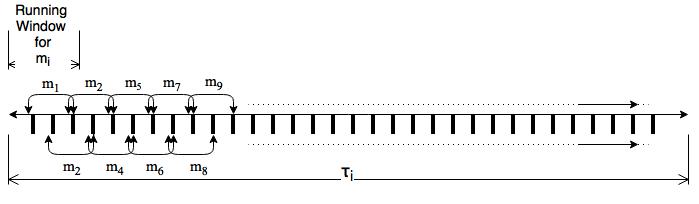}
\caption{Sliding windows for m in $\tau_i$}\label{fig:1a}		
	\end{subfigure}
%	\quad
	
	\begin{subfigure}[t]{1in}
		\centering
		\hspace*{-3cm}
		\includegraphics[scale=0.38]{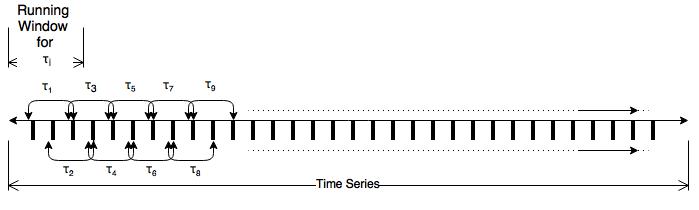}
		\caption{Sliding windows for $\tau_i$ in time series}\label{fig:1b}
	\end{subfigure}
	\caption{Showing the sliding windows m and $\tau_i$ in $\tau_i$ and time series respectively}\label{fig:1}
\end{figure}
%%%%%%%%%%%%%%%%%%%%%%%%%%%%%%%%%%%%%%%%%%%%%%%%%%%%%%%%%%%%%%%%%%%%%%%%%%%%%%%%%%%%%%%%%%%%%%%%%%%%%%%%%%%%%%%%%%%%%%%%%%%%%%

%%%%%%%%%%%%%%%%%%%%%%%%%%%%%%%%%%%%%%%%%%%%%%%%%%%%%%%%%%%%%%%%%%%%%%%%%%%%%%%%%%%%%%%%%%%%%%%%%%%%%%%%%%%%%%%%%%%%%%%%%%%%%%\section{Methodology}
For solving the mode mixing problem observed in the calculation of empirical mode functions(IMF) using Empirical Mode Decomposition we can follow the following algorithm.
\begin{enumerate}
\item Calculate the Permutation Entropy of the time series for a particular length($\tau$) with short sequence length(m) around each point.
\item Extract the high entropic portions from the time series and run EMD in each portion.
\item Calculate the PE values for each IMFs.
\item Compare the PE values of IMFs and the original time series and replace the high entropic portions with appropriate IMFs.
\item Now run EMD in both time series i.e, the repaired original data and the intermittent extracted component to get the desired result. 
\end{enumerate}

\subsection{Detailed Methodology}
As shown in the Fig.2 the intermittent portion of our signal can be easily detected with the help of Permutation Entropy Algorithm. The increase in complexity of the signal in the intermittent part can be easily observed by the higher entropic value of the region. For detecting the onset and offset of the noise in our original signal the following algorithm is applied
\begin{itemize}
\item Calculation of permutation entropy around each point of our time series.
\item For detecting the region of increased permutation entropy
\begin{itemize}
\item Taking the maxima envelop of Permutation Entropy.
\item Calculating the statistical mode of the maxima envelop of the permutation entropy.
\item Considering the portions of the Entropy curve above the mode level and extracting the data from the original time series at that particular portion. 
\end{itemize}
\end{itemize}
 This algorithm seems very specific for our problem but subsequently we will have to come up with a more general algorithm to extract the noisy part(high frequency part) from the time series.\par
 Fig.2,3,4,5 represents the entire methodology for extracting the intermittent part from the original signal.

\begin{figure}	
\centering

\centering
\includegraphics[scale=0.18]{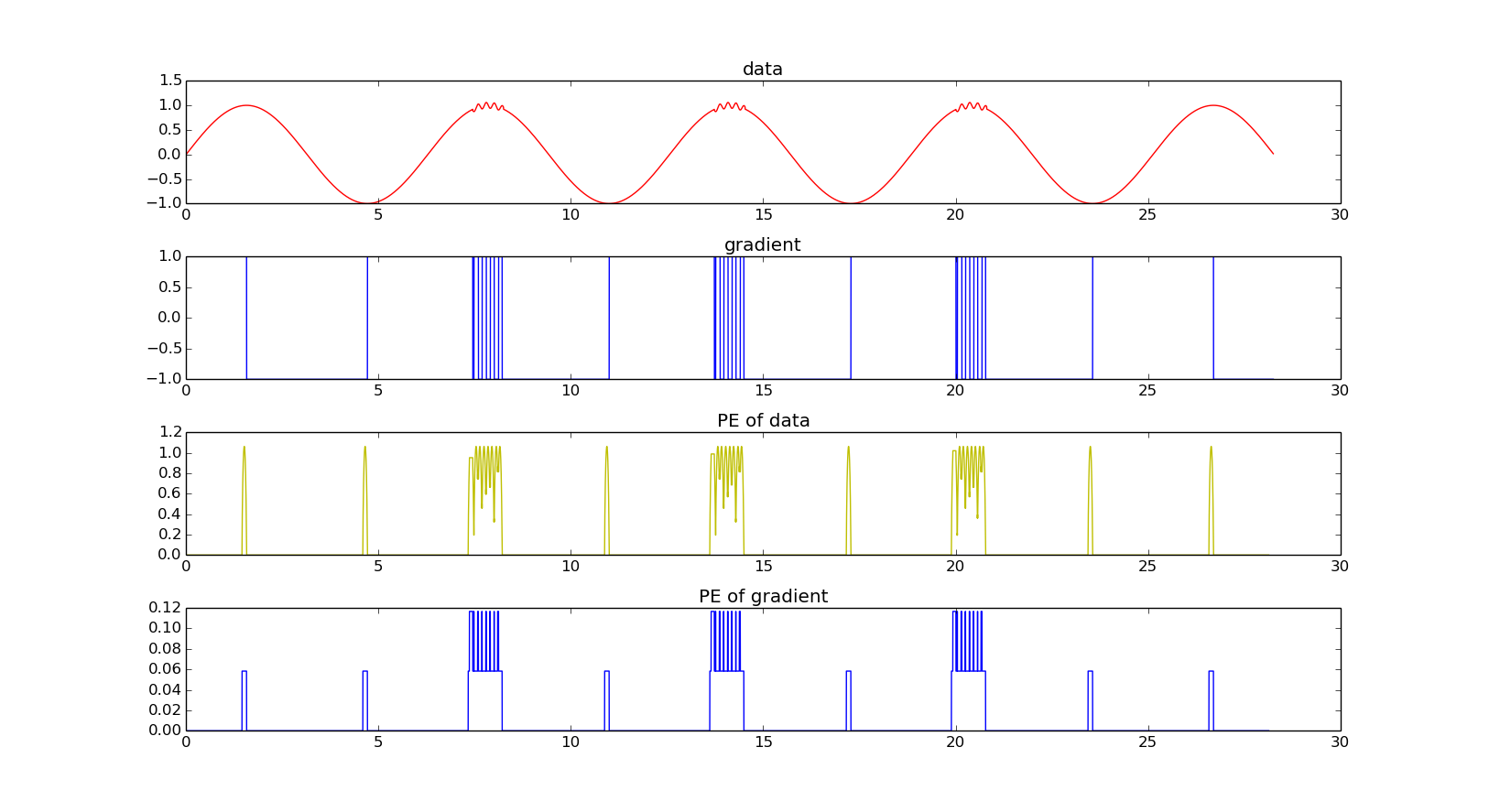}
\caption{Permutation Entropy for time series}
\label{fig:1b}

\centering
\includegraphics[scale=0.18]{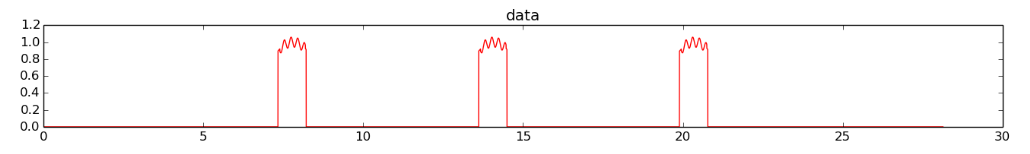}
\caption{Extracted time series based on high PE value}\label{fig:1a}		
\centering
\includegraphics[scale=0.18]{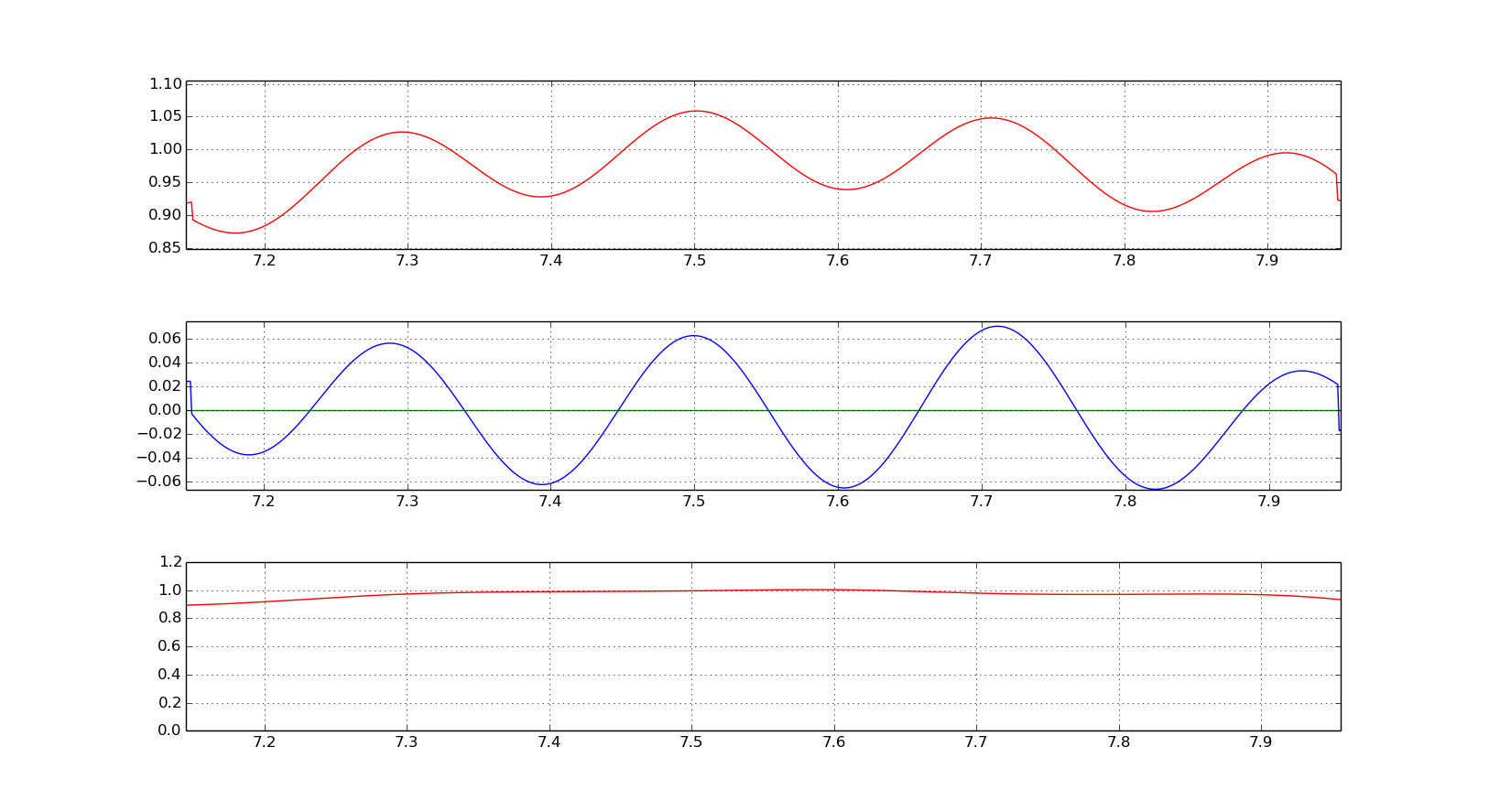}
\includegraphics[scale=0.18]{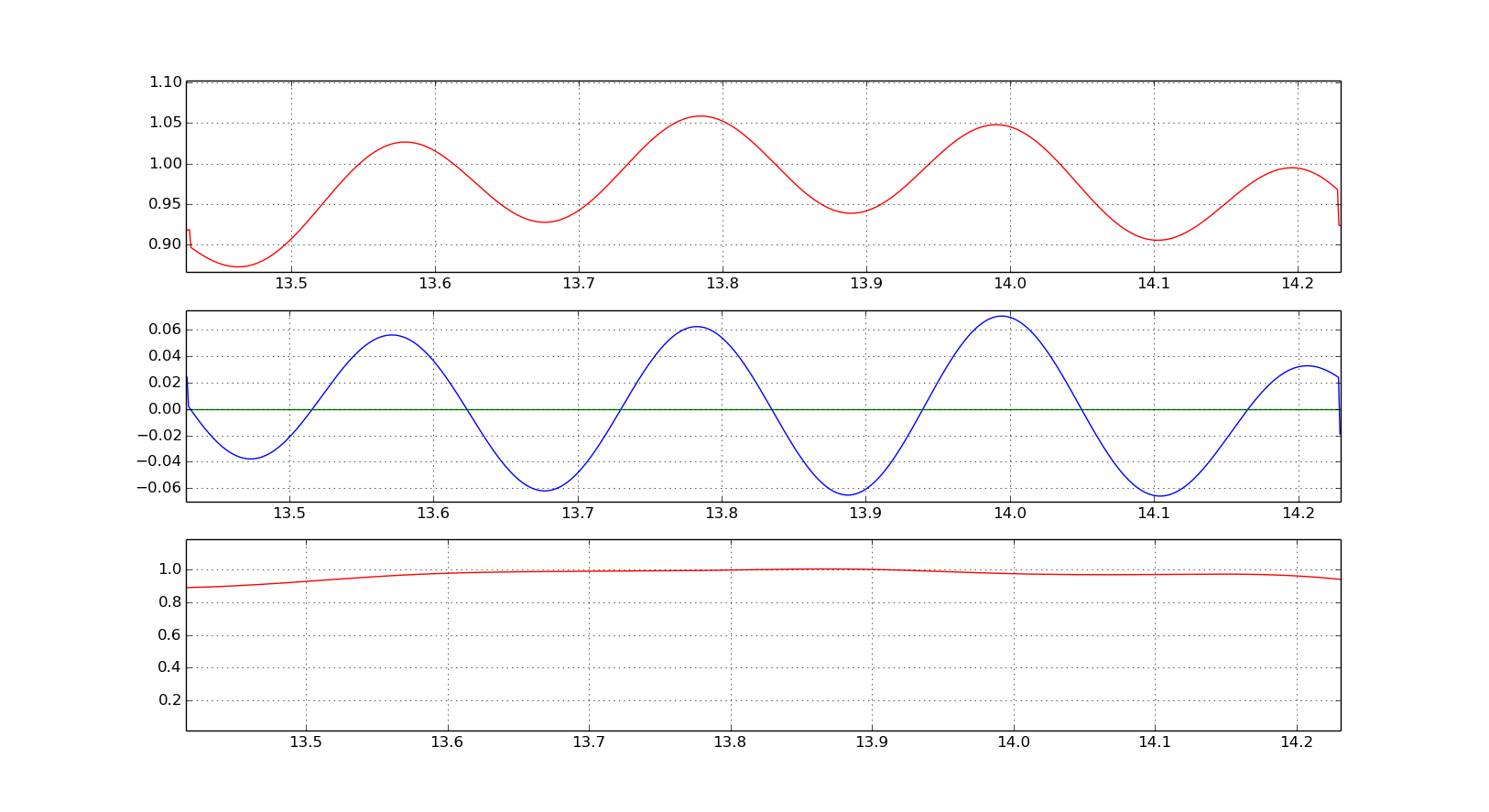}
\includegraphics[scale=0.18]{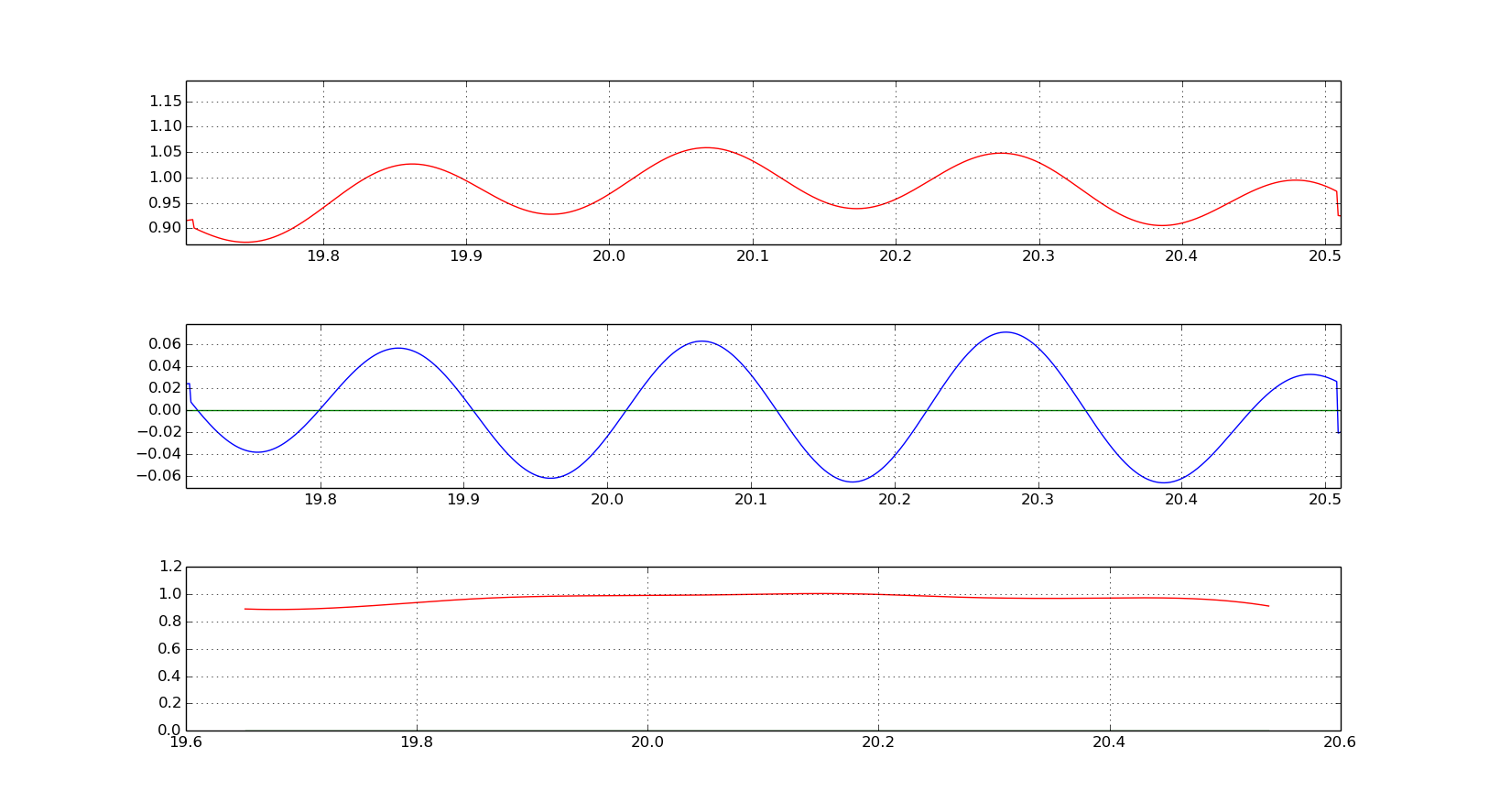}
\caption{Empirical Mode Decomposition for the selected time series portions}\label{fig:1b}
\centering
\includegraphics[scale=0.18]{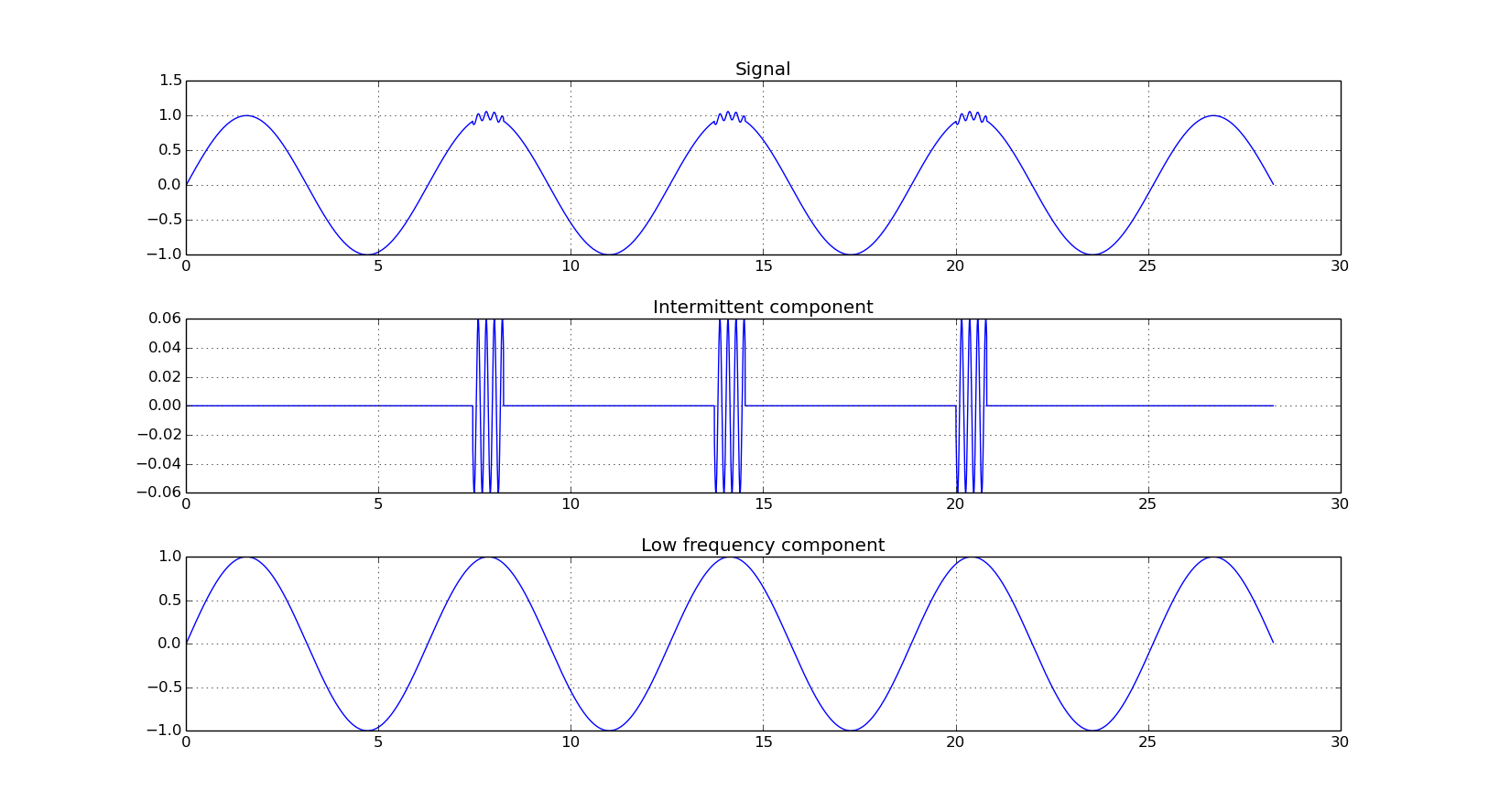}
\caption{Extracted Intermittent component and low frequency components}
\label{fig:1a}

\end{figure}
\subsection{Detecting the frequency variation}
 For our purpose we are particularly interested in detecting variation in frequency and want to avoid the detection of any amplitude change in our time series. But the above technique becomes problematic in certain cases where both amplitude as well as frequency changes are detected by the permutation entropy filter. So instead of running the PE directly on the time series it is better to run it on the gradient of the time series. According to the definition of gradient :
 $$grad(y)=\frac{\frac{dy}{dt}}{|{\frac{dy}{dt}}|} $$
	Where y and t are the data points and the time points in the time series. In this way every sequence of the time series gets a value of $+1$ and $-1$ for increasing and decreasing portions respectively. Afterwards applying PE algorithm on the gradient set gives a better result than the previous one.
%%%%%%%%%%%%%%%%%%%%%%%%%%%%%%%%%%%%%%%%%%%%%%%%%%%%%%%%%%%%%%%%%%%%%%%%%%%%%%%%%%%%%%%%%%%%%%%%%%%%%%%%%%%%%%%%%%%%%%%%%%%%%

%%%%%%%%%%%%%%%%%%%%%%%%%%%%%%%%%%%%%%%%%%%%%%%%%%%%%%%%%%%%%%%%%%%%%%%%%%%%%%%%%%%%%%%%%%%%%%%%%%%%%%%%%%%%%%%%%%%%%%%%%%%%%
\section{Conclusion}
 Even though various modification to EMD has been proposed like EEMD\cite{huang2} to resolve the mode mixing problem observed in IMF determination. But most of them seem computationally expensive and inappropriate for application in real time data analysis. Proposed algorithm describes the selective implementation of EMD/HHT which can adaptively determine the desired portions with the help of local entropic value and extract useful information.
 %%%%%%%%%%%%%%%%%%%%%%%%%%%%%%%%%%%%%%%%%%%%%%%%%%%%%%%%%%%%%%%%%%%%%%%%%%%%%%%%%%%%%%%%%%%%%%%%%%%%%%%%%%%%%%%%%%%%%%%%%%%%%

%%%%%%%%%%%%%%%%%%%%%%%%%%%%%%%%%%%%%%%%%%%%%%%%%%%%%%%%%%%%%%%%%%%%%%%%%%%%%%%%%%%%%%%%%%%%%%%%%%%%%%%%%%%%%%%%%%%%%%%%%%%%%%
\section{Acknowledgement}
The author would like to thank Department of Engineering Cybernetics,NTNU for the financial support through Jens Balchen's scholarship without which this work might not have been possible.
%%%%%%%%%%%%%%%%%%%%%%%%%%%%%%%%%%%%%%%%%%%%%%%%%%%%%%%%%%%%%%%%%%%%%%%%%%%%%%%%%%%%%%%%%%%%%%%%%%%%%%%%%%%%%%%%%%%%%%%%%%%%%%

%%%%%%%%%%%%%%%%%%%%%%%%%%%%%%%%%%%%%%%%%%%%%%%%%%%%%%%%%%%%%%%%%%%%%%%%%%%%%%%%%%%%%%%%%%%%%%%%%%%%%%%%%%%%%%%%%%%%%%%%%%%%%%

%%%%%%%%%%%%%%%%%%%%%%%%%%%%%%%%%%%%%%%%%%%%%%%%%%%%%%%%%%%%%%%%%%%%%%%%%%%%%%%%%%%%%%%%%%%%%%%%%%%%%%%%%%%%%%%%%%%%%%%%%%%%%%

\end{document}